\documentclass[runningheads]{cl2emult}

\usepackage{makeidx}  
\usepackage{graphicx} 
\usepackage{subeqnar} 
\usepackage{multicol} 
\usepackage{cropmark} 
\usepackage{eso}      
\makeindex            

\begin{document}
\newcommand\Msun {M_{\odot}\ } \def\ltsima{$\; \buildrel < \over \sim
\;$} \def\simlt{\lower.5ex\hbox{\ltsima}} \def\gtsima{$\; \buildrel >
\over \sim \;$} \def\simgt{\lower.5ex\hbox{\gtsima}}
\title*{Studying the Stellar Populations \protect\newline
of the Local Group with VLT }
\toctitle{Studying the Stellar Populations \protect\newline
of the Local Group with VLT}
%
%
\titlerunning{Stellar Populations with VLT}
%
\author{Eline Tolstoy}
\authorrunning{Eline Tolstoy}
%
%
\institute{ESO, 
Karl-Schwarzschild str. 2,
D-85748 Garching bei M\"{u}nchen, Germany
}

\maketitle              


\section{Introduction}
The best chance we have to understand star formation and how it
proceeds in the Universe is going to come from detailed studies
of the numerous different environments found within the Local
Group (LG). Present day star formation in our Galaxy occurs 
exclusively in metal rich environments (Z $\sim Z_{\odot}$), so
if we want to study how low metallicity stars form (and thus
understand observations of galaxies at high-redshift)
we have to look beyond our Galaxy, 
to the smallest star forming dwarf galaxies, which can
have extremely low metallicities (Z $\sim0.02-0.05 Z_{\odot}$).
Of course in its entirety a stellar population always contains
the complete details of the star formation history of a galaxy, however
this information is often hard to disentangle retroactively.
We also have much to learn from the Magellanic Clouds 
(Z $\sim 0.1-0.3 Z_{\odot}$), although because they are 
undergoing interactions with our Galaxy and each other their
evolutionary picture and its general applicability 
less obvious. 
In our LG there are also a number of ``remnants'', or galaxies which
which currently do not form stars 
({\it e.g.} the dSph, such as Carina, Leo~I, Ursa Minor, etc..).  
It is not straight forward to draw
parallels between galaxies which are forming stars and 
those which aren't. This is of course because star formation has
such a dramatic impact upon a galaxy, and 
alternative methods have to be used to make the most basic of comparisons
of properties ({\it e.g.} metallicity, mass, luminosity evolution). 
It is necessary to put all the dwarf galaxies into a global picture
if we are to draw
meaningful conclusions about their star formation properties
({\it e.g.} Ferrara \& Tolstoy 1999).
Many of the small LG galaxies contain direct
evidence of complicated star formation histories 
({\it e.g.} Smecker-Hane {\it et al.} 1994; Tolstoy {\it et al.} 1998; 
Gallart {\it et al.} 1999), which suggests that 
star formation patterns can change dramatically 
over long time scales. This kind of evolutionary behaviour can
have a dramatic impact upon the accurate
interpretation of galaxy redshift surveys (cf. Tolstoy 1998).

As in all scientific endeavors the most useful approach to solving
a complex question is to apply a number of different techniques, 
compare the results and then,  with the help of some theoretical
insight, form a consistent picture.
The most important
advances are likely to come not only from the ability to push fainter
and deeper into interesting regions of the Colour-Magnitude diagram
(CMD), but also from the ability to carry out detailed spectroscopic
analyses of individual stars in the same CMDs.
The first images and spectra
to come out of the Paranal Observatory from FORS1,
({\it e.g.} see Kudritski, this volume; Appenzeller, plenary talk), 
have offered the exciting two-fold promise: the ability to go very faint
and exquisite image quality.
This new era of the ground based 8-10m telescope promises to push
forward our understanding of galaxy evolution via the study of
nearby systems. 

\begin{figure}
\centering
\includegraphics[width=1.\textwidth]{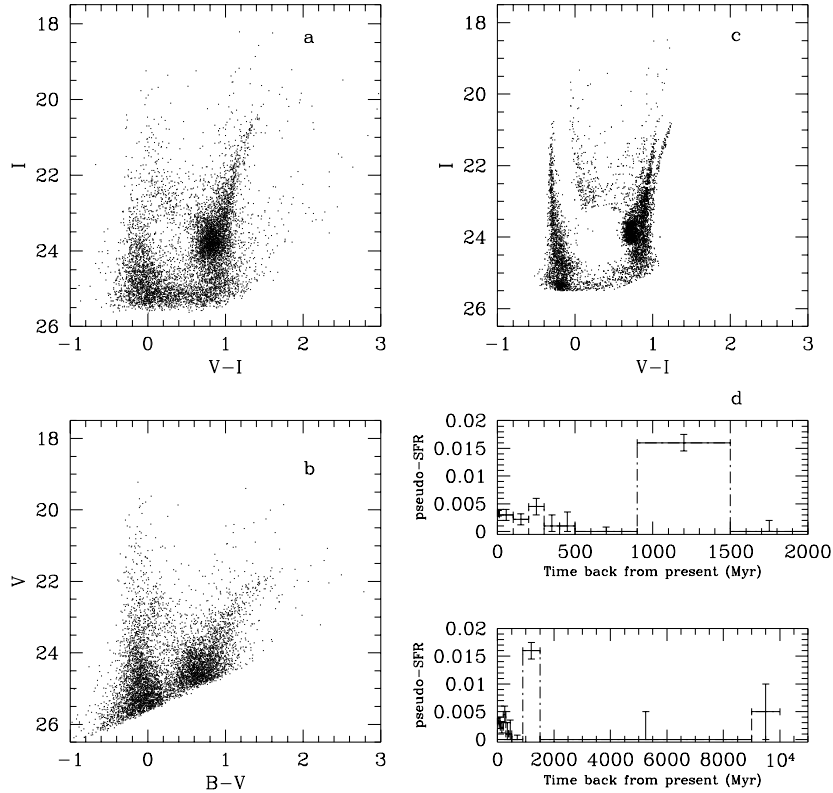}
\caption[]{Here are the results of the analysis of 
HST/WFPC2 data of nearby dwarf irregular galaxy
Leo~A (Tolstoy {\it et al.}  1998).  In a.  is the V$-$I, I 
CMD, 1 orbit exposure time per filter.  In b.  is the B$-$V, V CMD, 2 
orbits in B.  In c.  is the best match Monte-Carlo simulation model (in 
V$-$I, I) found for these data convolved with the theoretical measurement 
error distribution, and in d.  is the SFH that created the model CMD which 
best matches these data. See Tolstoy {\it et al.} (1998) for more details..}
\label{eps1}
\end{figure}

\section{Colour-Magnitude Diagrams}

The study of resolved stellar populations in CMDs as means of
understanding the properties of nearby stars and galaxies has a
distinguished historical tradition, going back to Leavitt almost a
century ago, and including the careers of 
Hubble and Baade ({\it e.g.} Baade 1963). It was
during the 1950's that the CMD provided the impetus to understand the,
then, holy grail of stellar evolution, using the newly operating
Palomar 200inch telescope, to observe Galactic star clusters.  
In the 1990s we have come so
far that stellar evolution theory is understood in sufficient detail
that we can make numerous predictions related to 
structures in a CMD ({\it e.g.} Bertelli {\it et al.} 1994; Tolstoy \& Saha 1996;
Girardi {\it et al.} 1996; Aparicio {\it et al.} 1996; Tolstoy 1999),
based upon variations of age and metallicity in star forming regions
many Gyr ago. Thus we have the ability to disentangle complex star
formation histories of all nearby galaxies. with sufficiently deep and
accurate CMDs.

Improved image quality is arguably of equal importance to an increase
in aperture size. Digging deep in to a CMD is only a worthwhile
exercise if the errors can be minimised.  Much of the
difficulty in interpreting CMDs comes from distinguishing effects
which come from photometric
errors due to image crowding and
those caused by the star
formation properties of a galaxy. 
The {\it Hubble Space Telescope} (HST) has been leading 
the way in providing
high quality CMDs ({\it e.g.} Dohm-Palmer {\it et al.} 1997, 1998; Gallagher {\it et al.} 1998;
Tolstoy {\it et al.} 1998; Cole {\it et al.} 1999; Gallart {\it et al.} 1999; 
see Figure~1), but the
collecting area and instruments available on VLT will be important for
further progress. In the case of low luminosity nearby galaxies 
HST images are completely uncrowded, and cover
only a fraction of a galaxy, so using VLT can be an equally good option, 
given an average seeing of $\simlt$ 0.6arcsec ({\it e.g.} see Fig~2).
The resulting CMDs (Fig.~3) show a wealth of detail, and a much
narrower red giant branch than previous observations. We also unequivocally
define the presence blue main sequence stars in Antlia, with an age in the
range 1-2 Gyr. 

\begin{figure}
\centering
\includegraphics[width=.5\textwidth]{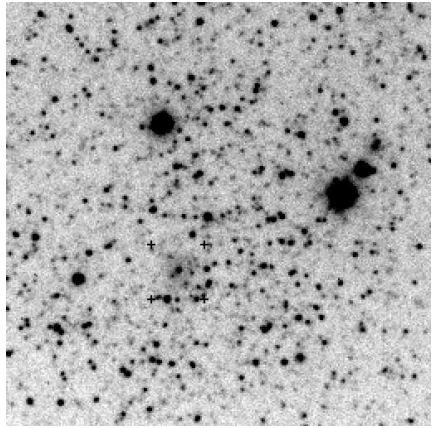}
\caption[]{
This is the central section (50arcsec squared) of the combined I filter
FORS1 images (total, 5400sec) of the 
Antlia dwarf galaxy taken under excellent seeing
conditions on 30th and 31st January 1999. The average seeing was below
0.5~arcsec for the entire integration period. It can be seen that in this
image crowding is not a problem, and the integrations could clearly be even
longer.
One point of note, is that an unusual source (in between the crosses here),
remarked upon by Sarajedini {\it et al.} (1997), 
it is visible in the I band (I$\sim$22.5) and
not in the V (V$<$25). It is clearly a
background galaxy in the FORS1 image. 

}
\label{eps2}
\end{figure}
\begin{figure}
\centering
\vskip-5.5cm
\includegraphics[width=1.\textwidth]{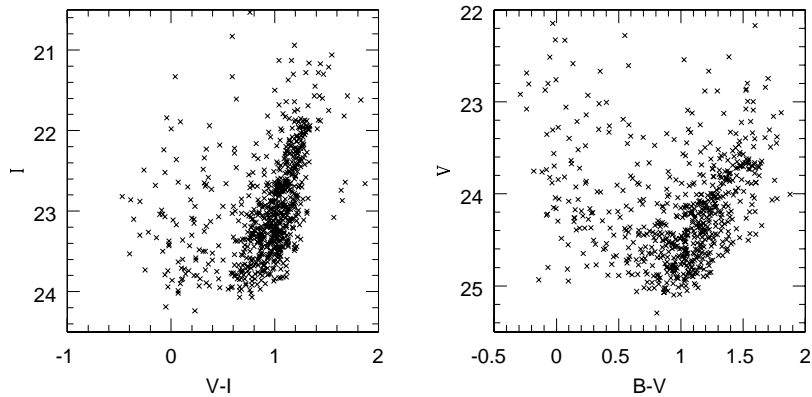}
\caption[]{
Here we show the first CMDs from the FORS1 science verification data
of the Antlia dwarf galaxy. They include stars within a circle of 3.3arcmin
diameter centred on Antlia. The larger this area the more extensive will
be the foreground contamination, but stars from Antlia
are seen out to the edges of the FORS1 frame (7~arcmin across).
The total integration times are 2400sec in B and V and 5400sec in I.
There is a clear ``blue plume'' of stars, which is real. Inspection of
the images shows that they are 
neither background galaxy confusion
nor a cosmetic problem on the CCD. They could represent
a comparatively young population of stars in Antlia (between 1 and 2 Gyr
old). There is no evidence for extensive reddening internal to Antlia,
but the calibrations used here are still preliminary.
The (preliminary) distance from the tip of the red giant branch is
1.4~Mpc, slightly further away than previous estimates.
}
\vskip -2cm
\label{eps3}
\end{figure}

\section{Metallicity Evolution}

The global effects of stellar evolution upon the
interstellar medium of a galaxy are surprisingly poorly
understood, especially in metal poor environments. 
It is critical to our understanding of galaxy evolution to
determine how the fraction of metals in a galaxy build up over time
by the star formation process. This
information is preserved in stellar abundance patterns of a
galaxy, however
distinguishing the effects of age and metallicity is often a complex task.
Purely photometric determinations from CMD analysis 
cannot provide unique solutions because of 
age-metallicity degeneracy on the red giant branch.
It is necessary to directly
measure the abundance of a large sample stars 
to understand how the metallicity has
Changed with time ({\it e.g.} Edvardsson {\it et al.} 1993, in our Galaxy).
If we can add independent
metallicity information about individual stars to CMD analysis, 
then we can 
determine much more accurate star formation histories.

There are several observational
approaches to this problem, all of which 
benefit from intermediate and high resolution spectrographs on
large telescopes ({\it e.g.} HIRES, FORS1/2, UVES). 
The large aperture of VLT or Keck is crucial
to overcome the limitation of the readout noise of the detectors, which is  
the limiting factor on smaller telescopes.

\begin{itemize}

\item
Metallicity evolution can be predicted from 
HII region emission line diagnostics with some 
success 
({\it e.g.} Matteucci \& Tosi 1985; Kobulnicky \& Skillman 1996; 
Pagel \& Tautvai\u{s}ien\.{e} 1998), 
although this is not independent of uncertain model assumptions.
There are also a number of LG galaxies with very faint HII regions about
which we know very little about the basic abundance measures.

\item
The direct relation between the Ca~II triplet absorption
index and stellar metallicities in metal poor systems, as defined by
Da Costa \& Armandroff (1995), affords a convenient and independent way
to measure the relative metallicities of evolved 
red giant stars, which cover the age range 
1$-$10~Gyr. This has been successfully applied to stellar populations
in the Magellanic Clouds ({\it e.g.} 
Olszewski {\it et al.}1991; Da Costa \& Hatzidimitriou 1998). 

\item
In the case of the brighter giants in nearest by galaxies, higher resolution
spectroscopy is possible for an accurate analysis of abundance
patterns [{\it e.g.} McCarthy {\it et al.} 1995 (M31); 
Shetrone, Bolte \& Stetson 1998 (Draco); 
Venn 1999 (SMC)]. These detailed
abundance analyses provide an more accurate picture of how chemical
enrichment occurs in different environments. For example, 
the pattern of element enrichment ({\it e.g.} C, N, O, Na, Mg and Al to Fe) 
in globular cluster giants is found to differ from field Population~II 
giants in the Galaxy halo (Pilachowski, Sneden \& Kraft 1996; 
Shetrone 1996).

\end{itemize}

\section{Summary}

The power of high
quality images and large collecting area of the VLT is impressive, and
there are clearly many exciting discoveries waiting to be made. 
We hope for further steps forward from the study of resolved
stellar populations with VLT, as from Palomar before it, because now
we will be able to probe a much large range of
star forming environment within the LG, and specifically very low metallicity
environments.
The tremendous gains in image quality and collecting area
now available with the VLT on Paranal make it 
important to survey the 
resolved stellar populations of all the accessible 
galaxies in our LG. This will provide a uniform picture of the 
properties of stellar populations
of galaxies with a wide variety of mass, metallicity, gas
content etc.
This has begun with FORS1 science verification 
observations in January 1999, with
BVI images of the LG galaxy Antlia (see Fig.~3).
The
study of individual stars and star-formation regions
in the nearby universe is the only way we will understand the
observations of galaxy populations at high redshift.

\subsubsection{Acknowledgments:} 
The Science Verification data used in this paper are public, and
were not presented at the Antofagasta meeting (only as press release
images by Prof. I. Appenzeller in his plenary talk), the 
analysis here was made after the data were made public in May 1999.
For more information, look at the ESO web page: 
{\tt http://www.eso.org/science/ut1sv/}

\clearpage
\addcontentsline{toc}{section}{Index}
\flushbottom
\printindex

\end{document}